\documentclass[preprint,12pt,3p]{elsarticle}

\usepackage{amssymb}
\usepackage[table]{xcolor}
\usepackage{multirow}
\usepackage{bigstrut}
\usepackage{makecell}
\usepackage{booktabs}
\everypar{\looseness=-1}

\usepackage[font=small,skip=0pt]{caption}
\begin{document}

\begin{frontmatter}

\title{How practical is it? Machine Learning for Identifying Conceptual Interoperability Constraints in API Documents}
\author[label1]{Hadil Abukwaik\corref{cor1}}
\address[label1]{University of Kaiserslautern, 67663 Kaiserslautern, Germany}
\address[label2]{IESE Fraunhofer, 67663 Kaiserslautern, Germany}

\cortext[cor1]{I am the corresponding author. URL: abukwaik.com}
\ead{abukwaik@cs.uni-kl.de}

\author[label1]{Mohammed Abufouda}
\ead{abufouda@cs.uni-kl.de}

\author[label1]{Thejashree Nair}
\ead{thejunair@gmail.com}

\author[label1,label2]{Dieter Rombach}
\ead{rombach@cs.uni-kl.de}

\date{\today}

\begin{abstract}
Building meaningful interoperation with external software units requires performing the conceptual interoperability analysis that starts with identifying the conceptual interoperability constraints of each software unit, then it compares the systems' constraints to detect their conceptual mismatch. We call the conceptual interoperability constraints (the COINs) that can be of different types including structure, dynamic, and quality. Missing such constraints may lead to unexpected mismatches, expensive resolution, and running-late projects. However, it is a challenging task for software architects and analysts to manually analyze the unstructured text in API documents to identify the COINs. Not only it is a tedious and time-consuming task, but also it needs knowledge about the constraint types. In this article, we present and evaluate our idea of utilizing machine learning techniques in automating the COIN identification, which is the first step of conceptual interoperability analysis, from human text in API documents. Our empirical research started with a multiple-case study to build the ground truth dataset, on which we contributed our machine learning COIN-Classification Model. We show the model's robustness through experiments using different machine learning text-classification algorithms. The experiments' results revealed that our model can achieve up to $87\%$ accuracy in automatically identifying the COINs in text. Thus, we implemented a tool that embeds our model to demonstrate its practical value in industrial context. Then, we evaluated the practitioners' acceptance for the tool and found that they significantly agreed on its usefulness and ease of use. 
\end{abstract}


\end{frontmatter}


\section{Introduction}
\label{sec1}

In the past, a software system was developed by a single organization to provide a tightly focused support for certain tasks with specific purposes. Today, software providers are urged to adopt integration solutions of independent software systems built by different organizations~\cite{barns1991making, boehm1999cots}. However, to build a meaningful interoperation between separately developed software systems, performing thorough conceptual interoperability analysis is required. In our research, this analysis consists of two steps. The \textit{first step} is to identify the conceptual interoperability constraints of the intended to interoperate systems. We define the \textbf{Co}nceptual \textbf{i}nteroperability co\textbf{n}straints (\textbf{COINs}) as the characteristics that control the meaningful exchange of data and functions between systems. We classify these constraints into the following classes: Syntax, Semantics, Structure, Dynamics, Context, and Quality~\cite{HA2}. For example, a contextual COIN restricts an offered sport-related software functionality to process data from elderly users. The \textit{second step} is to map the systems' constraints to detect their conceptual mismatches and decide on the feasibility and cost of resolving them. 

Accordingly, integrating black-box software units like Web Service or Platform APIs requires software architects and analysts to start with analyzing the shared API documents to identify the COINs in order to enable finding the existing mismatches before implementing the integration. Otherwise, unexpected conceptual mismatches can prevent the whole interoperation or make its results meaningless or harmful. This consequently causes resolution expenses at later stages of integration projects~\cite{garlan}. For example, a contextual mismatch can completely hinder a desired interoperation with the previously mentioned sport-related functionality, when the integration requires serving users from different age groups. 

Current analysis approaches rely on manual investigation of shared API documents~\cite{halle}. However, manual reading and inspection of natural language text in these documents to find the constraints is an exhausting, time-consuming, and error-prone task~\cite{wu}. Also, it requires knowledge about the various conceptual constraint types along with linguistic analysis skills. 

In this article, we extend our proposed idea of utilizing machine learning techniques to identify the COINs in the text of API documents (published in~\cite{HA7} as a component of our analysis framework~\cite{HA2}). That is, we focus on supporting the first step of the conceptual interoperability analysis. In particular, we expand the previous work with further experiments that show enhanced results through tuning the ML algorithms' parameters. Besides, we show a new technical contribution, in which we developed a software tool that facilitates practical benefits of our ML-based ideas. Moreover, we present an empirical evaluation study, in which we evaluate the tool with practitioners from the software industry.

Our research goal is to support software architects and analysts in performing effective and efficient identification of COINs in API documentations by employing machine learning (ML) techniques in a practical way. We followed a systematic empirical-based research methodology, which has three parts. \textit{In the first part}, we conducted a multiple-case study that yielded our first contribution, which is a ground truth dataset. This dataset is a community-reusable asset in the form of a repository of textual sentences that we collected from multiple API documents and manually labeled them with a specific COIN class. \textit{In the second part}, we contributed a classification model for the COINs using the ground truth dataset, and we evaluated it through experiments using different ML text-classification algorithms. Our experiments revealed promising results towards automating the identification of COINs in text of API documents. We achieved $70.4\%$ precision and $70.2\%$ recall for identifying seven classes of constraints (i.e., Syntax, Semantics, Structure, Dynamics, Context, Quality, and Not-COIN). This increased to $81.9\%$ precision and $82.0\%$ recall for identifying two classes (i.e., COIN, Not-COIN). In follow-up experiments, we tuned the algorithms' parameters, which increased the recall and precision of the two-class identification by $5\%$. \textit{In the third part}, we developed a tool to demonstrate the value of our ideas in serving software architects during their interoperability analysis task. The tool allows architects to select sentences from API document webpages, and it checks and reports the existence of COINs along with their types. Then, we empirically evaluated the tool with 51 practitioners through a survey after trying the tool. The survey results reveal a significant agreement on the tool usefulness and ease of use in extracting the COINs from text of API documents. 

The rest of the article is organized as follows. Section 2 introduces a background, Section 3 overviews the related works, and Section 4 outlines our research methodology. Section 5, 6, and 7 detail our three research parts. Section 8 includes the summary and future work.
\vspace*{-5mm}
\section{Background}
\label{sec2}
\subsection{Conceptual Interoperability Constraints}
\label{subsec2.1}
The presented work in this article is based on the Conceptual Interoperability Constraints (COIN) Model~\cite{HA2}, which focuses on conceptual and architectural restrictions of interoperable software systems and can be applied to different system types (e.g., information systems, embedded systems, mobile systems, etc.). As COINs govern the software system’s interoperability with other systems, missing or wrongly understanding them may defect the desired interoperability by leading to conceptual inconsistencies or meaningless results. There are six classes of COINs that we summarize as the following: (1) \textit{Syntax COINs} that state the constraints packaging (e.g., used terminology or modeling language). (2) \textit{Semantic COINs} that express meaning-related constraints (e.g., goals of methods). (3) \textit{Structure COINs} that depict the system’s elements, their relations, and arrangements affecting the interoperation results (e.g., data distribution). (4) \textit{Dynamic COINs} that restrict the behavior of interoperating elements (e.g., synchronization feature). (5) \textit{Context COINs} that pertain to external settings of the interoperation (e.g., user and usage properties). (6) \textit{Quality COINs} that capture quality characteristics related to exchanged data and services (e.g., response time).
\vspace*{-2mm}
\subsection{Machine Learning for Text Classification}
\label{subsec2.2}
To automate the COIN detection in text, we employed well-known ML algorithms used in text-classification problems (e.g., Na{\"\i}veBayes~\cite{nb} and Support Vector Machine (SVM)~\cite{svm}). Algorithm accuracy depends on the size and quality of the dataset~\cite{banko} that contains manually labeled sentences with one of the predefined classes. The classification process includes:\\
- \textit{Building the classification model}, in which all features of the sentences in the dataset are identified and modeled mathematically. In our research, we used popular techniques for building our model: (1) Bag-of-Words (BOWs)~\cite{chu} that considers each word in a sentence as a feature, and accordingly a document is represented as a matrix of weighted values. We chose it as it is simple and requires no manual tagging for the sentences or manual creation for classification rules. (2) N-Grams~\cite{ngram} that considers each N adjacent words in a sentence as a feature, where $(N \textgreater \hspace{0.05cm} 0)$.\\
- \textit{Evaluating the classification model}, in which the manually labeled dataset is divided into a training and testing sets. The training set is used for training the ML classification algorithm on the features captured in the model, while the testing set is used for evaluating the classification accuracy. For our research, we used \textit{k-fold Cross-validation}~\cite{cross}, in which our ground truth dataset (i.e., COINs Corpus) is divided into k folds. The $(k-1)$ folds are used for training and one fold for testing. Finally, an average of $k$ evaluation rounds is computed.
The ML binary (or two-class) classification algorithms are all about finding the best classifier model that represents the distribution from which the data comes and that separates the two classes effectively. Finding this most appropriate model not only requires finding the most proper ML algorithm, but also it requires a computationally expensive tuning for the algorithm's parameters. This is called \textit{optimization}, which improves classification results by choosing the best combination of values for the different algorithm parameters. This can be implemented using, for example, the \textit{Grid Search} method~\cite{hsu2003practical}, which is an exhaustive search process to find the best values from all possible combinations of the tuned parameters. Another option to do the parameter tuning is to adopt a randomized search~\cite{bergstra2012random}. 
\vspace*{-5mm}
\section{Related Work}
\label{sec3}
A number of previous works proposed automating the identification of some constraints from API documents. Wu et al.~\cite{wu} targeted parameters’ dependency constraints, Pandita et al.~\cite{pandita} inferred formal specifications for methods’ pre/post conditions, and Zhong et al.~\cite{zhong} recognized resource specifications. We complement these works and elaborate on Abukwaik et al.~\cite{HA2} idea of extracting a comprehensive set of conceptual interoperability constraints. 

On a broader scope, other works proposed retrieving information to assist software architects in different tasks. Anvaari and Zimmermann~\cite{anvaari} retrieved architectural knowledge from documents for architectural guidance purposes. Figueiredo et al.~\cite{figueiredo} and Lopez et al.~\cite{lopez} searched for architectural knowledge in emails, meeting notes, and wikis for proper documentation purposes. Although these are important achievements, they do not meet our goal of assisting architects in interoperability analysis tasks.

In general, our work and the aforementioned related works intersect in the utilization of natural language processing techniques in retrieving specific kind of information from documents. However, they used rule-based and ontology-based retrieval approaches, while we explored ML classification algorithms that are helpful for information retrieval in natural language text. Add to this, our systematic research contributed a reusable ground truth dataset for \textit{all COIN types} that enables related research replication and results' comparison.
 
\section{Research Methodology}
\label{sec4}
In this research, we systematically revealed the potentials of automating the COINs extraction from API documents using ML techniques. Our research goal formulated in terms of GQM template~\cite{GQM} is: \textit{to} support conceptual interoperability analysis \textit{for the purpose of} improvement \textit{with respect to} effectiveness and efficiency in identifying the COINs \textit{from the viewpoint of} software architects and analysts \textit{in the context of} analyzing text in API documents within software integration projects. We translate the goal into research questions:\\
\textbf{RQ1}: \textit{What are the existing conceptual interoperability constraints, COINs, in the text of API documentation?} This question explores the current state of COINs in real API documents. It also aims at building the ground truth dataset (i.e. COINs Corpus representing a repository of sentences labeled with their COIN class). This forms a main building block towards the envisioned automatic extraction idea.\\
\textbf{RQ2}: \textit{How effective and efficient would it be to use ML techniques in automating the extraction of COINs from text in API documentations?} This question explores the actual benefits of utilizing ML in supporting architects and analysts in analyzing the text. It results in our classification model that will be evaluated through well-known ML classification algorithms.\\
\textbf{RQ3}: \textit{To what extent would our ML-based software tool be accepted by practitioners in automating the extraction of COINs from text in API documentation?} This question explores the practitioners' acceptance for our idea of utilizing the ML capabilities to support them in analyzing the conceptual constraints in API documents. That is, it finds the perceived usefulness and ease-of-use of our tool-supported method in industry.

In order to achieve the stated goal and answer the aforementioned questions, we performed our research in three main parts as follows:\\
\textbf{Research Part 1 (Multiple-case study)}. We systematically explored the state of COINs in six cases of API documentation. This resulted in a ground truth dataset (i.e., COINs Corpus). We detail the study design and results in Section 5.\\
\textbf{Research Part 2 (Experiments)}. We used the ground truth dataset to build the COIN-Classification Model, which we validated its accuracy using different ML classification algorithms. We detail the experiments process and results in Section 6.\\
\textbf{Research Part 3 (Tool development and its empirical evaluation)}. We developed a software tool that embeds our COIN-Classification Model, then we evaluated its acceptance by allowing a sample of practitioners to try it and report their experience about it through a survey. We describe the tool, its evaluation study, and results in Section 7.

Our systematic research provides traceability between the different activities and their results and enables researchers to independently replicate our work and compare results.

\section{Multiple-Case Study: Building the Ground Truth Dataset for COINs}
\label{sec5}
\subsection{Study Design}
\label{subsec5.1}
\textbf{Study goal}. We aim at answering the first research question RQ1 that we stated in Section 4. In order to do so, we needed to examine real-world API documentations to discover the state of conceptual interoperability constraints in them.

\textbf{Research method}. We performed a multiple-case study with literal replication of cases from different domains in order to collect significant evidence and draw generalizable results.

\textbf{Case selection criteria}. For systematic selection of cases of API documentations, we considered the following selection criteria:\\
\textit{SC1: Mashup Score}. This is a published statistical value\footnote{Programmable web: http://www.programmableweb.com/apis/directory} for the popularity of a Web Service API in terms of its integration frequency into new bigger APIs.\\
\textit{SC2: API Type}. This can be either Web Service API or Platform API.\\
\textit{SC3: API Domain}. This is the application domain for the considered API document (e.g., social blogging, audio, software development, etc.).

\textbf{Analysis unit}. Our case study has a holistic design, which means that we have a single unit of analysis. This unit is ``the sentences in API documents that include COIN instances''. To document and maintain the analyzed sentences, we designed a data extraction sheet that we implemented as an MS Excel sheet. This sheet consists of demographic fields (i.e., API name, date of retrieval, mashup score, API type, API domain, and no. of sentences) and analysis fields (i.e., case id, sentence id, sentence textual value, and the COIN class).

\textbf{Study protocol}. Our multiple-case study protocol includes three main activities that are adapted from the process proposed by Runeson~\cite{runeson}. The study activities are case selection, case execution, and cross-case analysis, which we detail in the next subsection. 

\subsection{Study Execution and Results}
\label{subsec5.2}
\textbf{Case selection}. Based on our predefined case selection criteria, in August 2015 we chose six API documents. Four documents were from the Web Services type (i.e., SoundCloud, GoogleMaps, Skype, and Instagram) and two from the Platform type (i.e., AppleWatch and Eclipse-Plugin Developer Guide). The cases cover different application domains (i.e., social micro-blogging, geographical location, telecommunication, social audio, and software development environment). Regarding the mashup criteria, the four Web Service API documents cover a wide range of scores (i.e., from 30 for Skype and to 2582 for GoogleMaps).

\textbf{Data preparation}. For each case, we started with fetching the API documentation for the selected case from its online website. Then, we read the documents and determined the webpages that had textual content offering conceptual software description and constraints (e.g., the Overview, Introduction, Developer Guide, API Reference, Summary, etc.). Subsequently, we started processing the text in chosen webpages by performing the following:\\
- \textit{Automatic filtering}. We implemented a simple PHP code using Simple HTML DOM Parser\footnote{Simple HTML DOM: http://simplehtmldom.sourceforge.net/} library to filter out the text noise (i.e., headers, images, tags, symbols, HTML code, and JavaScript code). Thus, we passed the URL link of the chosen webpage (input) to our implemented code to get back a .txt file containing the textual content of the webpage (output). This filtered out information might be input for other explorative studies seeking knowledge from sources other than the human text that we consider the scope of this study. \\
- \textit{Manual filtering}. The automatic filtering falls short in excluding specific types of noise (e.g., text and code mixture, references like ``see also'', copyrights, etc.). This noise could mislead the machine learning in our later research steps, so we removed them manually.

\textbf{Data collection}. For each case, we cut the content of the text file resulted from the previous step into single sentences within our designed data extraction sheet (.xls file) that we described in Subsection 5.1. Afterwards, we filled all the fields of the data sheet for each sentence except for the ``COIN class'' field that we filled within the next step. \textit{Note that}, we maintained a data storage, in which we stored the original HTML webpages of the selected API documentations, their text file, and their excel sheet (see the study webpage\footnote{http://abukwaik.com/site/multiple-case-study\\ Password:\texttt{jss\_2017} (After publication, it will be available under a sharing agreement)}). This enables later replication of our work by researchers as API documents changes frequently.

\textbf{Data analysis}. For each case, we manually analyzed each collected sentence in the extraction sheet and carefully assigned it a COIN class. This classification was based on an interpretation criteria, which is the COIN Model with its six classes (i.e., Syntax, Semantic, Structure, Dynamic, Context, and Quality). We added a seventh class for sentences with no COIN instance (i.e., Not-COIN class). For example, the sentence ``A user is encapsulated by a read-only Person object.'' was classified as a ``Structure COIN'', while, ``You can also use our Sharing Kits for Windows, OS X, Android or iOS applications'' was classified as a ``Not-COIN'' as it did not express a conceptual constraint, but rather a technical information.

Obviously, the case execution process consumed time and mental effort, especially in the data analysis step. Table~\ref{tab:caseshare} summarizes the distribution of our collected 2283 sentences among the cases along with the effort that we spent in executing them. Noticeably, SoundCloud and Instagram have small documents, and consequently, they have the smallest share of sentences included in our study (i.e., $9.5\%$ and $11\%$). Meanwhile, Eclipse documentation is the largest and consequently has the highest share of sentences (i.e., $28.5\%$).

\begin{table*}[htbp]
\footnotesize
  \centering
  \caption{Case-share of sentences and execution effort.}
    \begin{tabular}{|c|c|c|}
    \hline
     \rowcolor[rgb]{ .851,  .851,  .851}\textbf{API Document} & \textbf{Number of sentences} & \textbf{Execution efforts  (Hours)} \\
    \hline
    {Sound Cloud} & 219   & 7.7 \\
    \hline
    {GoogleMaps} & 473   & 6.5 \\
    \hline
    {AppleWatch} & 360   & 8 \\
    \hline
    {Eclipse Plugin } & 651   & 12 \\
    \hline
    {Skype} & 325   & 4.5 \\
    \hline
    {Instagram} & 255   & 4.8 \\
    \hline
    \rowcolor[rgb]{ .851,  .851,  .851}\textbf{Total} & \textbf{2283} & \textbf{43.5} \\
    \hline
    \end{tabular}%
  \label{tab:caseshare}%
\end{table*}%

\textbf{Cross-Case Analysis (Answering RQ1: What are the existing types of conceptual interoperability constraints, COINs, in the text of current API documents?)}. After executing all cases, we arranged the incrementally classified sets of sentences from all cases (i.e., 2283 sentences) into one repository that we call the \textit{ground truth dataset} or \textit{the COINs Corpus} as called in ML. We developed two versions of this dataset as follows:

- \textit{Seven-COIN Corpus}, in which, each sentence belongs to exactly one of the seven classes (i.e., Not-COIN, Dynamic, Semantic, Syntax, Structure, Context, or Quality). That is, we have no sentence in the corpus that has more than one class, as our classes are orthogonal.  

- \textit{Two-COIN Corpus}, in which, each sentence belongs to one of two classes rather than seven (i.e., COIN or Not-COIN). In fact, the Two-COIN Corpus is derived from the Seven-COIN Corpus by abstracting the six COIN classes into one class. Table~\ref{tab:exampleofcontent} shows the difference between the two Corpora with example sentences.

The aim of building the two corpus versions is to better investigate the performance results of the ML algorithms in the later research experiments (details are in Section 6).
\begin{table*}[htbp]
\footnotesize
  \centering
  \caption{Example of content in the Seven-COIN and Two-COIN Corpus.}
    \begin{tabular}{|c|p{10cm}|c|c|}
    \hline
    \rowcolor[rgb]{ .851,  .851,  .851} \textbf{ID} & \multicolumn{1}{c|}{\textbf{Sentence}} & \textbf{Seven-COINs} & \textbf{Two-COINs} \\
    \hline
    {s1} & You can also use our Sharing Kits for Windows, OS X, Android or iOS applications. & \cellcolor[rgb]{ .949,  .949,  .949} Not-COIN & \cellcolor[rgb]{ .949,  .949,  .949} Not-COIN \\
    \hline
    {s2} & When it is finished manipulating the object, it releases the lock. & \cellcolor[rgb]{ .949,  .949,  .949} Dynamic & \cellcolor[rgb]{ .949,  .949,  .949} COIN \\
    \hline
    {s3} & A user is encapsulated by a read-only Person object. & \cellcolor[rgb]{ .949,  .949,  .949} Structure & \cellcolor[rgb]{ .949,  .949,  .949} COIN \\
    \hline
    {s4} & A user’s presence is a collection of information about the users’ availability, their current activity, and their personal note. & \cellcolor[rgb]{ .949,  .949,  .949} Syntax & \cellcolor[rgb]{ .949,  .949,  .949} COIN \\
    \hline
    {s5} & A dynamic notification interface lets you provide a more enriched notification experience for the user. & \cellcolor[rgb]{ .949,  .949,  .949} Semantic & \cellcolor[rgb]{ .949,  .949,  .949} COIN \\
    \hline
    {s6} & This service is not designed to respond in real time to user input. & \cellcolor[rgb]{ .949,  .949,  .949} Context & \cellcolor[rgb]{ .949,  .949,  .949} COIN \\
    \hline
    {s7} & Your interfaces need to display information quickly and facilitate fast navigation and interactions. & \cellcolor[rgb]{ .949,  .949,  .949} Quality & \cellcolor[rgb]{ .949,  .949,  .949} COIN \\
    \hline
    \end{tabular}%
  \label{tab:exampleofcontent}%
\end{table*}%

\textit{COIN-share in the contributed ground truth dataset}. The Not-COIN class, which expresses technical constraints rather than conceptual ones, is dominant among the other six classes (i.e., $42\%$). The Dynamic and Semantic classes have the second and third biggest shares. Remarkably, the Structure, Syntax, Quality, and Context instances are very few with convergent shares ranging between $1\%$ and $5\%$ of the dataset.

\textit{COIN-share in the cases}. On a finer level, we have investigated the state of COINs in each case rather than in the whole ground truth dataset. We found that the content of each API document was focused on the Not-COIN, Dynamic and Semantic classes similarly as in the aggregated findings on the complete dataset. For example, in the case of AppleWatch documentation, $40.8\%$ of the content is for Not-COIN, $26.1\%$ for Dynamic, and $25\%$ for Semantic. Add to this, all cases had less than $10\%$ of its content to the Structure, Syntax, Quality, and Context classes (e.g., in the Eclipse-Plugin case this is only $8.5\%$). 

\subsection{Discussion}
\label{subsec5.3}
\textbf{Technical-oriented API documentations}. The Not-COIN class reserves $42\%$ of the sentences in the investigated parts of the API documents that were supposed to be conceptual (i.e., overview and introduction sections). A noteworthy example is the GoogleMaps case, which took it to an extreme level of focus on the technical information (i.e., $63\%$ of its content was under the Not-COIN class, $11.2\%$ for Dynamic class, $13.1\%$ for Semantic class, and the rest was shared by the other classes). Accordingly, it is important to raise a flag about the lack of sufficient information about the conceptual aspects of interoperable software units or APIs (e.g., usage context, terminology definitions, quality attributes, etc.). This needs to be brought to the notice of researchers and practitioners who care about the usefulness and adequacy of content in API documentations. This obviously has an influence on architects' effectiveness in conceptual interoperability analysis related activities.

\textbf{Considerable presence of Dynamic and Semantic constraints}. Our study findings reveal that the Dynamic and Semantic classes have apparently big shares in current API documents (i.e., $25\%$ and $24\%$ of the dataset). This reflects the favorable awareness about the importance of proper and explicit documenting of the API semantics (e.g., data meaning, service goal, conceptual input and output, etc.) and dynamics (e.g., interaction protocol, flow of data, pre- and post- conditions, etc.). Nevertheless, based on the tedious work we went through our manual analysis for the six cases, we believe that it would be of great help for architects and analysts to have clear borders between these two classes of constraints within the verbose of text. For example, it would be easier to skim the text, if the API goal gets separated from its interaction protocol, rather than blending them into long paragraphs. This would offer a better reading experience and consequently enhance the analysis results.

\textbf{COIN-deficiency in Platform and Web Service API documents}. From our investigated cases, we perceived a convention on assigning insignificant shares for the Structure, Syntax, Quality, and Context classes. Interestingly, the cases varied with regards to what they chose to slightly cover out of these four classes.

On one hand, the cases of Web Service APIs were the main contributors to the Context, Quality, and Syntax classes in the ground truth dataset. That is, the documents of GoogleMaps, SoundCloud, Skype, and Instagram provided $82.5\%$ of the Syntax COINs, $70.4\%$ of the Quality COINs, and $92\%$ of the Context COINs. Such a contribution cannot be related to the nature of Web Service APIs, as Platform ones need also to share these COINs explicitly. For example, it is critical for a FarmerWatch application to know the offered response time by the Notification service of AppleWatch APIs.

On the other hand, the Platform API documents participated with $56.1\%$ of the Structure COINs in the ground truth dataset, while the Web Service API documents participated with $\%43.9$. Note that, this is not related to the larger amount of sentences that these two documents contributed to the dataset, but rather due to the internal case share of Structure COINs. On average, the Platform API documents allocate about $6\%$ of their content to structural constraints, while Web Service API documents allocate about $3.6\%$ for them.

\textbf{Observed patterns for the dominant classes in the ground truth dataset}. From the considerable amount of sentences for the Not-COIN, Semantic, and Dynamic classes, we observed a number of patterns as frequently occurring terms and sentences. Of the \textit{Not-COIN class patterns}, which we found in $30.7\%$ of its instances, were the ``Technical Keywords'' that abbreviate software technologies (e.g., XML, iOS, XPath, etc.). Also, sentences starting with specific terms (e.g., ``for example'', ``for more information'', ``see'', etc.) recurred in $12.8\%$ of the Not-COIN instances. Meanwhile, of the \textit{Dynamic class patterns}, which recurred in $35.8\%$ of its instances, were terms related to actions and data/process flow that we gathered into a list called the ``Action Verbs'' (e.g., create, request, access, lock, etc.). Furthermore, $24\%$ of the Dynamic COIN sentences contained a conditional statement expressing a pre- or post- condition. Similarly, of the \textit{Semantic class patterns}, which were found in $18.8\%$ and in $16.4\%$ of its instances, were what we call ``Input/Output Terms'' (e.g., return, receive, send, etc.) and ``Goal Terms'' (e.g., allow, enable, permit, etc.).

We envision that using these patterns in combination with the Bag-of-Words (BOWs) in future experiments would enhance the results of the automatic COIN identification. 

\subsection{Threats to Validity}
\label{subsec5.4}
\textbf{Case bias}. Due to approach sensitivity to the quality and the quantity of sentences in the document case, we included multiple cases in building the ground truth that had a prominent role in our research. This also facilitates obtaining significant results and drawing generalizable conclusions. We literally replicated six cases (i.e., SoundCloud, GoogleMaps, Skype, Instagram, AppleWatch and Eclipse-Plugin Developer Guide) from two API types (i.e., Web Service and Platform).

\textbf{Completeness}. Due to resource limitations (i.e., time and manpower), we were unable to analyze the large API documents completely. However, we were careful with respect to selecting inclusive parts of such large documents. For example, out of the huge document of Eclipse APIs, we covered the Plugin part.

\textbf{Researcher bias}. Building the ground truth was very critical for our envisioned automatic COIN extraction idea. Therefore, to ensure building it in a way that guarantees results' accuracy and impartiality, we replicated the data analysis of each case (i.e., the manual classification of each sentence) by two researchers independently and based on the COINs Model as an interpretation criteria. In multiple discussion sessions, the researchers compared their classification decisions and resolved conflicts based on consensus.

\section{Experiments: Automatic Identification of COINs Using Machine Learning}
\label{sec6}

\subsection{Experiments Design}
\label{subsec6.1}
\textbf{Experiments goal}. We aim at answering the second research question RQ2 stated in Section 4. To do so, we needed to examine ML techniques to discover their potentials in automatically identifying the COINs in the text of API documents

\textbf{Research method}. We built a classification model and ran multiple experiments employing different ML text-classification algorithms. This method enables comparing the algorithms’ results and drawing solid conclusions about the ML advantages in addressing the challenges of manual COIN extraction from text in API documents.

\textbf{Evaluation method and metrics}. We used k-fold Cross-validation, which we explained in the background section, with $k = 10$. For evaluation metrics of classification accuracy, we used the following commonly used measures~\cite{powers}: (1) \textit{Precision}: the ratio of correctly classified sentences by the classification algorithm to the total number of sentences it classifies. (2) \textit{Recall}: the ratio of correctly classified sentences by the classification algorithm to the total number of sentences in the corpus. (3) \textit{F-measure}: the harmonic mean of precision and recall that is calculated as: $(2*Precision*Recall)/(Precision+Recall)$.

\begin{figure}
\centering
\includegraphics[height=5cm]{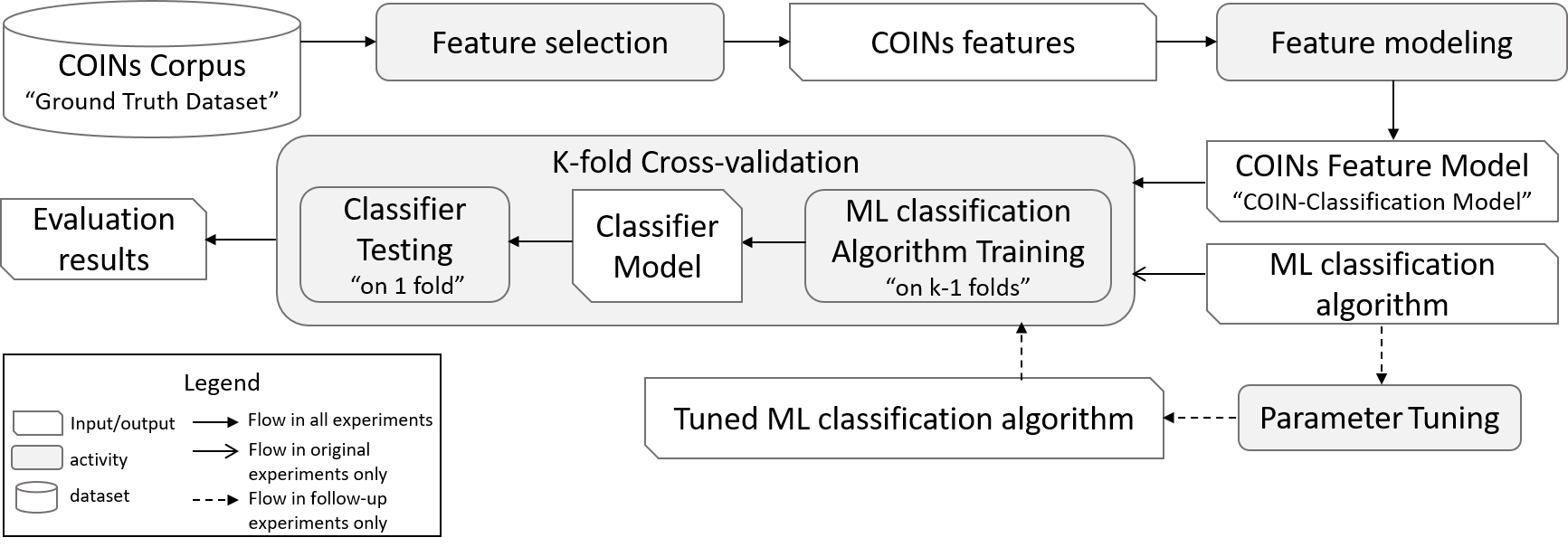}
\caption{ML experiments process.}
\label{fig:experimentproocol}
\end{figure}

\textbf{Experiments protocol}. Our experiments protocol includes three main activities that are: feature selection, feature modeling, and ML algorithms evaluation. We illustrate this protocol in Fig.~\ref{fig:experimentproocol}, and we describe it in details within the next subsection. We ran this protocol twice, once for the Seven-COIN Corpus and another for the Two-COIN Corpus. Also, we performed follow-up experiments, in which we ran the same protocol with automatic tuning for the parameters of the evaluated algorithms to see if we would get better results.
\vspace*{-3mm}
\subsection{Experiments Execution and Results}
\label{subsec6.2}
We performed our experiments on Weka v3.7.11\footnote{Weka: http://www.cs.waikato.ac.nz/ml/weka}, which is a suite of ML algorithms written in Java with result visualization capabilities. The execution started with processing the textual sentences in our contributed dataset (i.e., COINs Corpus) using natural language processing (NLP) techniques. The processing included tokenizing sentences into words, lowering cases, eliminating noise words (e.g., is, are, in, of, this, etc.), and stemming words into their root format (e.g., encapsulating and encapsulated are returned as encapsulate).

\textbf{Feature selection}. After processing the text, we identified the most representative features or keywords for the COIN classes within the COINs Corpus using the Bag-of-Words (BOWs) and N-Gram approaches, which we explained in the background section. That is, each sentence was represented as a collection of words. Then, each single word and each n-combination of words in the sentence were considered as features, where N was between 1 and 3. For example, in a sentence like ``A user is encapsulated by a read-only Person object'', the word ``encapsulate'' and the combination ``read-only'' were considered as two of its features. The output of this step was a set of features for the COINs Corpus.

\textbf{Feature modeling (Building the ML COIN-Classification Model)}. We transformed the whole COINs Corpus into a mathematical model. That is, we represented it as a matrix, in which headers contained all extracted features from the previous phase, while each row represented a sentence of the corpus. Then, we weighted each matrix cell [row, column] with the weight of the feature in the specific sentence. For weighting, we used the Term Frequency-Inverse Document Frequency (TF-IDF)~\cite{Robertson}, which is often used in text retrieval. This resulted in the COINs Feature Model (or the COIN-Classification Model), which is a reusable asset reserving knowledge about conceptual constraints in the API documents.

\textbf{COIN-Classification Model Evaluation (K-fold Cross-Validation, K = 10)}. We selected a number of well-known ML text classification algorithms (e.g., Na{\"\i}ve Bayes versions, Support Vector Machine, K-Nearest Neighbor KNN, and more). Then, we cross-validated the model using each algorithm in two steps. The first was a supervised training where the ML algorithm got to learn k-1 folds of our classification model with their COIN classes for k rounds. This resulted in our classifier model. The second step was testing the resulted classifier model on its prediction for the COIN class for 1 fold for k rounds.

\textbf{Evaluation results (Answering RQ2: How effective and efficient would it be to use ML in automating the extraction of COINs from text in API documents?)}.

\textit{Effectiveness of identifying the COINs using ML Algorithms}. The effectiveness results, with the default configuration of the algorithms' parameters, in terms of accuracy are:\\
- \textit{Seven-COIN Corpus Case}. The evaluation results showed that the best accuracy in automatically identifying seven classes of interoperability constraints in text was achieved by the ComplementNa{\"\i}veBayes algorithm (see Table~\ref{tab:algorithmsresult}). It achieved $70.4\%$ precision, $70.2\%$ recall, and $70\%$ F-measure. In the second place came Na{\"\i}veBayesMutinomialupdatable algorithm with about $5\%$ less accuracy than the former algorithm. The other algorithms had lower accuracy with the worst results from the KNN algorithms.\\ 
- \textit{Two-COIN Corpus Case}. By applying the same algorithms on the Two-COIN Corpus, we obtained better results. In particular, the accuracy increased with almost $11\%$ compared to the results in the Seven-COIN case with the ComplementNa{\"\i}veBayes algorithm. That is, the precision increased to $81.9\%$, recall to $82.0\%$, and F-measure to $81.9\%$. Similar to the previous case, Na{\"\i}veBayesMutinomialupdatable came in the second rank and the 2-Nearest Neighbor algorithm had the worst results as seen in Table 3. Note, we have achieved an improvement in accuracy compared to our preliminary investigation results~\cite{HA6}, in which we had F-measure of $62.2\%$ using the Na{\"\i}veBayes algorithm.

\begin{table*}[htbp]
\footnotesize
  \centering
  \caption{COINs identification results using different ML algorithms.}
    \begin{tabular}{|c|c|c|c|c|}
    \hline
    \rowcolor[rgb]{ .949,  .949,  .949} \multirow{-0.2}[4]{*}{\textbf{ML Classification Algorithm}} & \multicolumn{2}{c|}{\textit{\textbf{Seven-COIN Corpus}}} & \multicolumn{2}{c|}{\textit{\textbf{Two-COIN Corpus}}} \\
\cline{2-5}    \rowcolor[rgb]{ .949,  .949,  .949}       & \textbf{Precision} & \textbf{Recall} & \textbf{Precision} & \textbf{Recall} \\
    \hline
    \rowcolor[rgb]{ .949,  .949,  .949} \textit{\textbf{ComplementNa{\"\i}veBayes}} & \cellcolor[rgb]{ 1,  1,  1} $70.40\%$ & \cellcolor[rgb]{ 1,  1,  1} $70.20\%$ & \cellcolor[rgb]{ 1,  1,  1} $81.90\%$ & \cellcolor[rgb]{ 1,  1,  1} $82.00\%$ \\
    \hline
    \rowcolor[rgb]{ .949,  .949,  .949} \textit{\textbf{Na{\"\i}veBayesMutinomialupdatable}} & \cellcolor[rgb]{ 1,  1,  1} $66.00\%$ & \cellcolor[rgb]{ 1,  1,  1} $65.10\%$ & \cellcolor[rgb]{ 1,  1,  1} $81.90\%$ & \cellcolor[rgb]{ 1,  1,  1} $82.00\%$ \\
    \hline
    \rowcolor[rgb]{ .949,  .949,  .949} \textit{\textbf{Support Vector Machine (SVM)}} & \cellcolor[rgb]{ 1,  1,  1} $59.30\%$ & \cellcolor[rgb]{ 1,  1,  1} $60.00\%$ & \cellcolor[rgb]{ 1,  1,  1} $75.70\%$ & \cellcolor[rgb]{ 1,  1,  1} $75.70\%$ \\
    \hline
    \rowcolor[rgb]{ .949,  .949,  .949} \textit{\textbf{Random Forest Tree}} & \cellcolor[rgb]{ 1,  1,  1} $60.40\%$ & \cellcolor[rgb]{ 1,  1,  1} $56.30\%$ & \cellcolor[rgb]{ 1,  1,  1} $73.70\%$ & \cellcolor[rgb]{ 1,  1,  1} $73.90\%$ \\
    \hline
    \rowcolor[rgb]{ .949,  .949,  .949} \textit{\textbf{Simple Logistic}} & \cellcolor[rgb]{ 1,  1,  1} $52.50\%$ & \cellcolor[rgb]{ 1,  1,  1} $54.40\%$ & \cellcolor[rgb]{ 1,  1,  1} $68.20\%$ & \cellcolor[rgb]{ 1,  1,  1} $68.40\%$ \\
    \hline
    \rowcolor[rgb]{ .949,  .949,  .949} \textit{\textbf{KNN  K=1}} & \cellcolor[rgb]{ 1,  1,  1} $54.80\%$ & \cellcolor[rgb]{ 1,  1,  1} $45.50\%$ & \cellcolor[rgb]{ 1,  1,  1} $64.20\%$ & \cellcolor[rgb]{ 1,  1,  1} $52.30\%$ \\
    \hline
    \rowcolor[rgb]{ .949,  .949,  .949} \textit{\textbf{KNN  K=2}} & \cellcolor[rgb]{ 1,  1,  1} $49.80\%$ & \cellcolor[rgb]{ 1,  1,  1} $36.10\%$ & \cellcolor[rgb]{ 1,  1,  1} $64.40\%$ & \cellcolor[rgb]{ 1,  1,  1} $48.70\%$ \\
    \hline
    \end{tabular}%
  \label{tab:algorithmsresult}%
\end{table*}%

\textit{Efficiency of identifying the COINs using ML Algorithms}. Obviously, the machine beats the human performance in terms of the spent time in analyzing the text. As we mentioned earlier, analyzing the documents costed us about 44 working hours, while, it took the machine way less time. For example, training and testing the Na{\"\i}veBayesMultinominalupdate took about 5 seconds on our complete corpus with 2283 sentences). This efficiency would enhance when using machines with faster and more powerful CPU (we ran the experiments on a machine with Intel core i5 460 M CPU with 2.5 GHZ speed).

\subsection{Follow-up Experiments with Parameter Tuning for  ML Algorithms}
\label{subsec6.3}
In this subsection, we further investigate the effectiveness of our COIN-Classification Model through follow-up experiments, where we optimized the results by tuning the parameters of the ML algorithms. That is, instead of using the defaults of WEKA as in the previous experiments that offered the baseline accuracy, we tuned the parameters to get accuracy maximization with error minimization. In particular, we applied the Grid Search method as our dataset is relatively small. The space of the parameters that were tuned was extensive. For example, for the SVM we tuned the following parameters: the penalty method (L1 or l2), the loss function, the penalty value, and the degree $d$ of the polynomial kernel of SVM algorithm~\cite{Chapelle2002}. For the SVM case, we tested the whole possible space of parameters, which yielded 756 experiments each of them with unique setting for the parameters. The same approach was used for all the other classifiers. Regarding the implementation, we used the scikit-learn python library~\cite{scikitlearn}. Note, the parameter tuning with grid searching is very time consuming. For the small dataset we have, we managed to finish the training in almost two days with the same machine specifications we mentioned in the previous subsection.

The results of the follow-up experiments showed no improvement for the classification effectiveness (i.e., accuracy) in the case of the Seven-COIN Corpus. However, we got noticeably higher effectiveness in the case of the Two-COIN Corpus compared to the results reported in the previous section. Table~\ref{tab:parapemtertuning} shows only the best performing algorithms with significant accuracy improvements achieved via the parameter tuning. As seen, the highest accuracy result (i.e., F-measure = $87\%$) was obtained from the Polynomial SVM algorithm~\cite{svm} with kernel degree = 3. This result is 5\% higher than the best result achieved without parameter tuning. The other best performing tuned algorithms had achieved almost the same accuracy as the best achieved without parameter tuning. That is, the Logistic Regression with L2 regularization level~\cite{l2}, Linear SVM~\cite{svm}, and Logistic Regression with Stochastic Gradient Descent (SGD)~\cite{sgd} achieved F-measure of $81.9\%$, $81\%$, and $79.5\%$ respectively.

\begin{table}[]
\centering
\footnotesize
\caption{The results of the classification after parameter tuning.}
\label{tab:parapemtertuning}
\begin{tabular}{|
>{\columncolor[HTML]{EFEFEF}}l |c|c|c|}
\hline
\cellcolor[HTML]{EFEFEF} & \multicolumn{3}{l|}{\cellcolor[HTML]{EFEFEF}\textit{\textbf{Two-COIN Corpus}}} \\ \cline{2-4} 
\multirow{-2}{*}{\cellcolor[HTML]{EFEFEF}\textbf{ML Classification Algorithm}} & \cellcolor[HTML]{EFEFEF}\textbf{Precision} & \cellcolor[HTML]{EFEFEF}\textbf{Recall}  & \cellcolor[HTML]{EFEFEF}\textbf{F-measure} \\ \hline
\textit{\textbf{SVM (Polynomial, d=3)}} & $87.00\%$ & $87.00\%$ & $87.00\%$\\ \hline
\textit{\textbf{LogisticRegression (L2)}} & $81.70\%$ & $82.00\%$ & $81.85\%$\\ \hline
\textit{\textbf{SVM (Linear)}} & $81.00\%$ & $81.00\%$ & $81.00\%$\\ \hline
\textit{\textbf{LogisticRegression (SGD)}}& $80.00\%$ & $79.00\%$ & $79.50\%$\\ \hline
\end{tabular}
\end{table}

\begin{figure}[h]
    \centering
\includegraphics [scale=0.7]{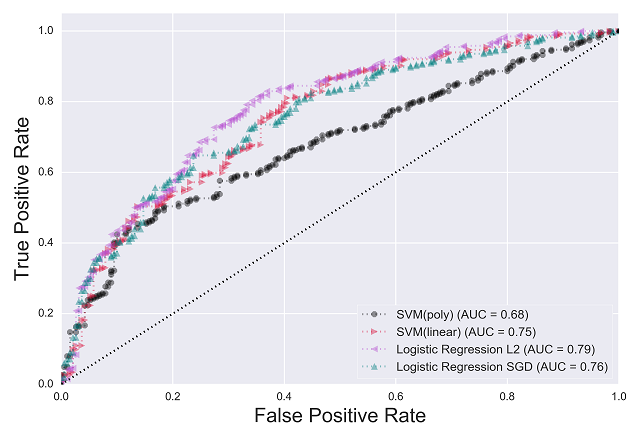}
    \caption{The area under FPR-TPR curve for classification algorithms after parameter tuning.}
    \label{fig:auc}
\end{figure}

While recall, precision, and F-measure inform us about the classification accuracy of the algorithms using our model, these measures do not take into account the true negatives~\cite{davis2006relationship}. Hence, we further investigated the False Positive Rate - True Positive Rate (FPR-TPR) curves for the different binary ML classification algorithms that we tuned their parameters (see Figure~\ref{fig:auc}). The curve of each algorithm is a plot of the trade-off between the algorithm ability to correctly detect the COINs (i.e., TPR or recall) and the number of incorrect alarms for COINs (i.e., FPR). Thus, the area under curve (AUC)~\cite{auc}, which has a value range from 0 to 1, measures how each algorithm is effective in segregating the two classes (i.e., COIN and Not-COIN). Therefore, the larger the AUC (or the closer the curve to the upper left corner), the higher the algorithm's probability in correctly classifying the sentences. 
\\Note that, the dashed diagonal line in the figure represents the curve for a random classification algorithm that has AUC of $0.5$. It is commonly used as a baseline to see if the other algorithms are useful. That is, an algorithm with AUC larger than $0.5$ is considered as non-random binary classification algorithm. Hence, the depicted algorithms in the figure show good classification effectiveness when compared to the random algorithm, which supports the validity of our COIN-Classification Model and the robustness of our automation idea.

\vspace*{-3mm}
\subsection{Discussion and Limitations}
\label{subsec6.4}
\textbf{Towards automatic conceptual interoperability analysis}. The achieved effectiveness in automatically identifying the constraints (i.e., $87\%$ F-measure) is promising and shows the potentials of our COIN-Classification Model in serving architects in the interoperability analysis tasks. We consider this accuracy high, as we compared the algorithms' results to our complete sentence-by-sentence manual analysis for the API documents, which we did for the sake of building a robust corpus. However, in practice, sentences are not examined in such a heavy way, especially when projects are limited in time and manpower. Hence, our model and its results are a step towards achieving a good level of intelligent automation for the classic software engineering practices that are error-prone and resource-consuming.

\textbf{Larger corpus, better accuracy results}. It is known in ML that the more classification classes you want to train the machine on, the more training data it requires to be fed with. This explains the higher accuracy we achieved using the Two-COIN Corpus compared to the Seven-COIN Corpus even with the same amount of sentences in both. Therefore, we plan to enlarge our corpus, to achieve better accuracy in identifying the seven COIN classes.

\textbf{Unbalanced amount of instances for each class in the corpus}. As noticed, the number of instances for the COIN classes is not balanced in the corpus. That is, dominant classes (i.e., Not-COIN, Dynamic, and Semantic) contribute with the majority of sentences in the data set (i.e., $91\%$). While, the other classes (i.e., Structure, Syntax, Quality and Context) are smaller and share the left $9\%$ of the corpus. This affects the classification accuracy of classes with fewer instances. Therefore, in future, we intend to increase the number of instances for these minor classes in the training data to achieve higher accuracy.

\textbf{Limited context}. The current version of our classification model cannot be used to identify COINs from any API document. This is due to the fact that our corpus is relatively small (i.e., ~3k sentences) and is built from six cases only (which has specific characteristics with regards to company size, document-writer features like his language or role, the maturity of the API, etc.). Hence, it will not be appropriate to generalize the features of the sentences in our small corpus to all existing sentences of all API documents. Thus, we intend in the future to enlarge the corpus (e.g., hundreds of thousands of sentences) to cover a wider range of API documents with different characteristics and to update the classification model with further features based on the new sentences. Accordingly, our tool support is currently reliable in the context of our six cases and their similar cases.
\vspace*{-5mm}
\section{COINer: An Empirically-Evaluated Supporting Tool}
\label{sec7} 

\subsection{Tool Description}
\label{subsec7.1}
\textbf{Goal}. To bring our idea to life and make it applicable in practice, we developed a novel tool, COINer, that facilitates the ML-based COIN extraction from text in API documents. Through its easy-to-use interfaces, it assists finding the conceptual constraints (i.e., the first step of the conceptual interoperability analysis) and consequently to understand their impact based on their class. Thus, the tool offers potential improvements for the next step of interoperability analysis (i.e., mismatch detection), especially for inexperienced analysts.

\textbf{Features overview}. The COINer tool is an add-on for Chrome Web Browser and it embeds our contributed ML COIN-Classification Model that we described in the previous section for the seven classes. The tool locates the COIN instances within the text of the webpage of an API document and shows their class in seconds. It also generates a separate report with all sentences that have COINs. Add to this, it allows architects to edit the automatically determined COIN class for a sentence and to send the feedback to the tool provider (install the tool and watch its demo on the study webpage.

\textbf{Detailed functional features with exemplary results}. Below we describe the functional features of our tool and we explain them with exemplary results on the SoundCloud API document case that we included within the multiple-case study.

\begin{figure}
\centering
\includegraphics[width=12cm, height=4.5cm]{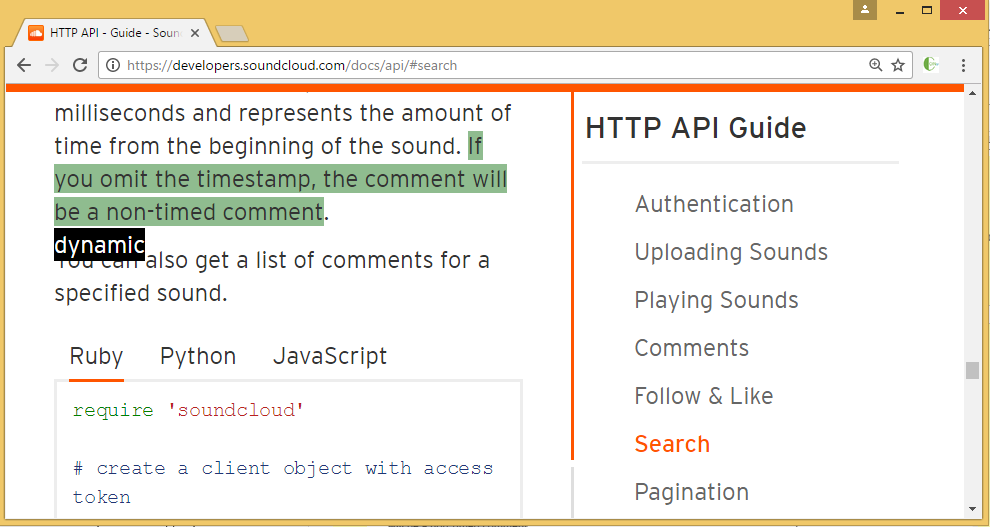}

\caption{A COINer example on automatic highlighting and annotation of a Dynamic COIN.}
\label{fig:coiner1}
\end{figure}

\textit{F1: Highlighting COINs within the text of the API document webpage}. This feature takes natural language sentences from the API documents as input and highlights the sentences that have COINs. By hovering on the highlighted sentence the user can see the COIN category (e.g., semantic, structure, etc.). The tool allows architects to select by the computer mouse either all text in the webpage of an API document or some text (i.e., a sentence or more). It also gives the user the opportunity to determine what COIN types to be highlighted (i.e., all COIN categories or some of them only). Fig.~\ref{fig:coiner1} shows an example of an automatically highlighted and annotated Dynamic COIN within the API document. Note, the different COIN categories are highlighted with different colors.

\begin{figure}
\centering
\includegraphics[width=12cm, height=5cm]{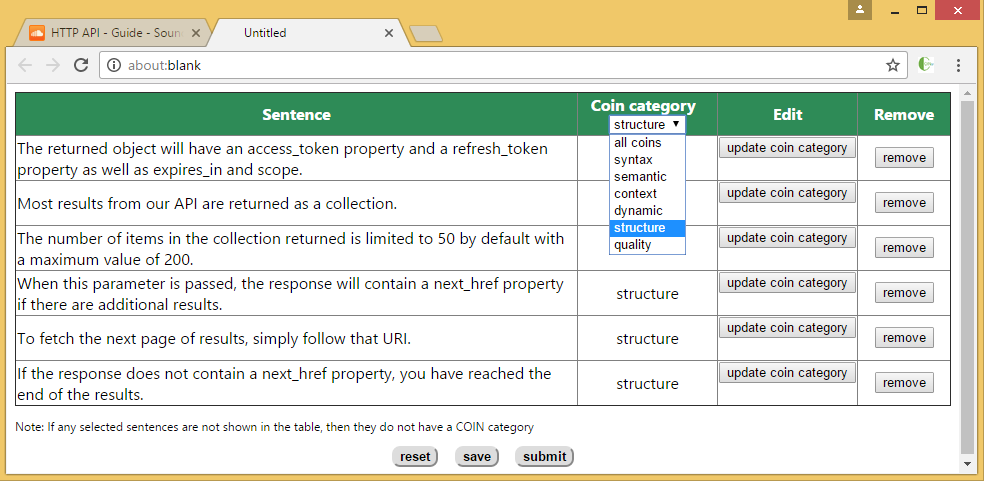}
\caption{A COINer example on a generated COINs report for Structure COINs.}
\label{fig:coiner2}
\end{figure}

\textit{F2: Generating a separate COINs report}. This feature takes natural language sentences from the API document as input and generates a separate report about the found COINs and their categories. As in F1, the tool offers architects to get the report either for the whole text or for some of it. It also gives the option to determine what COIN types to be displayed in the report, which can be printed or saved as an electronic file. Fig.~\ref{fig:coiner2} shows an example of a generated COINs report for Structure COINs in the SoundCloud document.

\textit{F3: Editing the COIN class for sentences and sending feedback}. As the highest achieved accuracy for our classifier is $\%81.9$, we give the users the possibility to update the COIN class for a sentence (using a drop-down list for the seven classes) for their own local copy of the generated report. The tool also offers the user to share his opinion with the tool providers through the ``submit'' button, then the update is saved in a special table for later use in maintaining the dataset and the model. Add to this, the users are given the option to edit their report copy by removing a COIN instance from the report based on their needs or opinions. The editing ``update'' and ``remove'' buttons are shown on the right side of Fig.~\ref{fig:coiner2} too. Also, the tool allows the user to reset the generated report to its original state through the ``reset'' button.  The ``submit'' and ``reset'' buttons are shown at the bottom of Fig.~\ref{fig:coiner2}.

\textbf{Architectural design}. The COINer tool has a client-server architecture, in which the server side has the major workload of the tool. As we plan to maintain and extend the tool (with larger corpus and accordingly with an enhanced classification model), we designed our tool to be modular (by separating concerns in modules interacting through well-defined interfaces). That is, the server side (back-end) contains the classification model, processes clients' requests, and classifies the sent text to be sent back to the client. While the client side (front-end) sends the users' requests to the server and renders the received output.

\textbf{Technical implementation}. We implemented the COINer tool as an add-on for the Chrome Web Browser. The tool encapsulates our contributed COIN-Classification Model and it mirrors its efficiency and accuracy that we described earlier in this subsection.

\textit{Development technology}. The server side of the tool is mainly implemented using Python and Java. It also contains the Apache Web Server and PHP scripts for handling the clients' requests. While the client side is implemented using Javascript and HTML code for presenting the tool functionalities and results to users. The communication between the client and the server sides is based on HTTP protocol and messages.

\textit{COINer working process}. The client-server communication starts by the client side sending the text that is desired to be classified to the server side. Consequently, the server starts with pre-processing the text using the NLP techniques, then it cuts it into separate sentences and sends them to the classifier model. The classifier responds with the COIN class for each sentence. Finally, the server aggregates the sentences with their classes and sends them back to the client side that represents it to the user in a proper format. 

\vspace*{-2mm}
\subsection{Tool Evaluation Study Design}
\label{subsec7.2}
\subsubsection{Study Design}
\label{subsec7.2.1}
\textbf{Study goal}.The main goal of this study formulated in terms of GQM template~\cite{GQM} is: \textit{to} analyze the implemented COINer tool \textit{for the purpose of} evaluation \textit{with respect to} its acceptance in terms of perceived usefulness and perceived ease-of-use \textit{from the viewpoint of} practitioners in industry \textit{in the context of} a survey. We translate this goal into the previously stated RQ3 in Section 4, which we split here into two sub-questions:\\
\textbf{RQ3.1}: \textit{To what extent do practitioners believe that using the COINer tool is useful in finding the conceptual interoperability constraints in the analyzed API document?}\\
\textbf{RQ3.2}: \textit{To what extent do practitioners believe that using the COINer tool is easy to use in finding the conceptual interoperability constraints in the analyzed API document?}

\textbf{Research method}.In order to achieve the stated goal and answer the aforementioned questions, we performed a survey through structured online questionnaire to systematically explore the practitioners' acceptance for the tool. Such a method aids in collecting evidence and drawing generalizable results about our tool.

\textbf{Target population}. We targeted software developers in industry who used API and read their documentations in their work. We invited about 62 developers from different software companies from different countries and got responses from 51 ($\%82.3$ response rate). The participants' experiences varied from low experience in API documents (less than 2 years) to very high experience (more than 10 years).

\textbf{Evaluation metrics}. To measure the acceptance of the tool, we used a subset of the metrics defined in the TAM model~\cite{davis1989user}. In particular, we derived the value of the perceived ease-of-use from the value of basic measures (i.e., perceived understandability, perceived flexibility, and ease-of-learning). While we derived the value of the perceived usefulness from the value of other basic measures (i.e., perceived results' completeness, perceived results' correctness, perceived tasks performance, and perceived task effort). The derived metrics are calculated as a median of the basic measures.

\textbf{Statistical hypotheses}. We derived the statistic null hypotheses (H0) and the corresponding alternative hypotheses (H1) from RQ3.1 and RQ3.2. Note, the arithmetic median of usefulness and ease-of-use are denoted by $\mu$ and it represents the answer on a response scale ranging from 1 (Strongly disagree) to 5 (Strongly agree). Thus, our hypotheses are:
\begin{center}\textit{H3.1,0  (Tool perceived usefulness)}: $\mu$ = 3 and H3.1,1 : $\mu$ $\textgreater$ 3.\\
\textit{H3.2,0  (Tool perceived ease-of-use)}: $\mu$ = 3 and H3.2,1 : $\mu$ $\textgreater$ 3.\end{center}

\textbf{Questionnaire and data collection}. The questionnaire included a background and tool demo video, questions on the tool perceived usefulness, questions on the tool perceived ease-of-use, and about the participant's experience about API documents. We implemented the questionnaire using Google Forms, which automatically saved the answers of each participant. The survey questionnaire is available at the previously provided study webpage.

\textbf{Data Analysis}. We analyzed the data using IBM SPSS Statistics 23. Our descriptive analysis includes the minimum, maximum, standard deviation, and median (see Table~\ref{tab:analysisresultofsurveydata}). We performed statistical analysis to explore how significantly different the ordinal answers are from a specific point using the One-Sample Wilcoxon signed-rank test, as the normality test showed that the collected data are not normally distributed.

\subsubsection{Execution and Results (Answering RQ3.1 and RQ3.2)}
\label{subsec7.2.2}
The study was executed in February 2017. The participants were given the survey link, which started with a video introducing the concepts and terminologies we use to ensure mutual understanding, and it also had a demo for the tool features. The participants were directed on installing the COINer tool and were asked to try it on some webpages for API documents that we had included in our multiple-case study (see Section 5). Then, each participant assessed the tool's usefulness and ease-of-use by answering our questionnaire.

By analyzing the collected data, we found that the surveyed practitioners had a significant agreement on considering the COINer tool as both useful and easy-of-use to identify the conceptual interoperability constraints in the text of API documents. Table~\ref{tab:analysisresultofsurveydata}, summarizes our descriptive and statistical analysis of the questionnaire results.

\begin{table}[]
\centering
\footnotesize
\caption{Analysis results of the survey data.}
\label{tab:analysisresultofsurveydata}
\begin{tabular}{ll|c|c|}
\cmidrule{3-4}
\multicolumn{2}{l|}{} & \cellcolor[HTML]{EFEFEF}Perceived Usefulness (H3.1) & \cellcolor[HTML]{EFEFEF}Perceived Ease-of-use (H3.2) \\ \hline
\multicolumn{2}{|l|}{\cellcolor[HTML]{EFEFEF}Min} & 2 & 1 \\ \hline
\multicolumn{2}{|l|}{\cellcolor[HTML]{EFEFEF}Max} & 5 & 5 \\ \hline
\multicolumn{2}{|l|}{\cellcolor[HTML]{EFEFEF}Median} & 4.25 & 4.12 \\ \hline
\multicolumn{2}{|l|}{\cellcolor[HTML]{EFEFEF}Standard deviation} & 0.76 & 0.77 \\ \hline
\multicolumn{1}{|c|}{\cellcolor[HTML]{EFEFEF}} & \cellcolor[HTML]{EFEFEF}Z & 1,103.00 & 997.5 \\ \cline{2-4} 
\multicolumn{1}{|c|}{\multirow{-2}{*}{\cellcolor[HTML]{EFEFEF}Test statistics $^a$}} & \cellcolor[HTML]{EFEFEF}p-value & 0.000* & 0.000* \\ \hline
\multicolumn{4}{|l|}{\cellcolor[HTML]{EFEFEF}$^a$ One-Sample Wilcoxon signed-rank test H0: Median(all respondent) = 3 (neutral); * p \textless 0.001} \\ \hline
\end{tabular}
\end{table}

According to the presented results, both null hypotheses (i.e., H3.1,0 and H3.2,0) are rejected with high statistical significance (i.e., p-value = 0). That is, there is a consensus from practitioners on accepting our COINer tool and they agree on its value in terms of its usefulness and ease-of-use during the conceptual interoperability analysis of API documents.
\vspace*{-7mm}
\subsubsection{Threats to Validity}
\label{subsec7.2.3}

\textit{Internal validity (Content validity)}. We peer-reviewed the design of the survey study with an expert researcher in empirical research. Further, we evaluated the survey in a pilot study with four participants to assess the understandability of the questionnaire.

\textit{Construct validity}. To avoid the risk of using improper measures, we used reliable metrics for acceptance (i.e., TAM) and we followed the GQM approach. We also had the study design peer reviewed by a second researcher with experience in empirical software engineering.

\textit{External validity (Representative sample)}. We got (N = 51) responses from software practitioners with different experiences about API documentation and from different organizations in different countries. Thus, we assume our results to be very likely representative for practitioners' acceptance of the tool as of February 2017. However, for better generalization and observation over time, further surveys with larger sample size are required.

\textit{External Validity (Completion rate)}. We ensured to have a small set of questions (i.e., 12 questions) in the survey questionnaire to increase the questions’ completion rate. Thus, we got a completely answered questionnaire by all the 51 respondents.
 \vspace*{-5mm}
\section{Summary and Future Work}
\label{sec8}
In this article, we have presented our extended ideas about supporting software architects in performing seamless conceptual interoperability analysis. The contribution pursued by this work was to utilize ML techniques for effective and efficient identification of conceptual interoperability constraints in text of API documents. Our research started with a multiple-case study that resulted in the ground truth dataset, which we used to build a ML COIN-Classification Model. Then, we started evaluating the practicality of our idea by evaluating our model in experiments using different ML algorithms. The results showed that the model allowed up to $70.0\%$ accuracy for identifying seven classes of interoperability constraints and $81.9\%$ for two classes. After tuning the parameters of the algorithms, the Two-COIN classification accuracy increased to $87\%$. We demonstrated the practical value of our ideas through a software tool that identifies the conceptual constraints in API documents and generates a COINs report. Finally, we conducted a survey with practitioners that revealed a significant agreement on the tool's usefulness and easy-to-use in the analysis task.

In the future, we plan to automate the manual filtering part of the data preparation and to analyze further API documents to advance the generalizability of our results. This would enrich the ground truth dataset as well, allowing better training and higher classification effectiveness for the ML algorithms. Also, we intend to further evaluate the actual effect of our ML-based identification of COINs on the final results of the whole conceptual interoperability analysis (i.e., on the detected systems' conceptual mismatches). This planned to be performed in industrial case studies and in a controlled experiment, in which a control group would perform the conceptual interoperability analysis manually in the traditional way, while the treatment group would have the support of ML through our COINer tool. 
\vspace*{-5mm}
\section*{Acknowledgement}
This work is supervised by Prof. Dieter Rombach and is funded by the Ph.D. Program of the CS Department and the Nachwuchsring of Kaiserslautern University.









\vspace*{-5mm}
\bibliographystyle{elsarticle-num}

\bibliography{reference}

\end{document}